# Study of the combined effect of temperature, pH and water activity on the radial growth rate of the white-rot basidiomycete *Physisporinus vitreus* by using a hyphal growth model

M. J. Fuhr[a,b], C. Stührk[a,b], M. Schubert[a], F. W. M. R. Schwarze[b] and H. J. Herrmann[a]

[a]*ETH Zurich, Institute for Building Materials, Computational Physics for Engineering Materials, Schafmattstrasse 6, HIF E18, CH-8093 Zurich, Switzerland*

[b]*EMPA, Swiss Federal Laboratories for Materials Science and Technology, Wood Laboratory, Group of Wood Protection and Biotechnology, Lerchenfeldstrasse 5, CH-9014 St Gallen, Switzerland*

*mfuhr@ethz.ch*

*Chris.Stuehrk@empa.ch*

*Mark.Schubert@empa.ch*

*Francis.Schwarze@empa.ch*

*hans@ifb.baug.ethz.ch*

Corresponding author: M.J. Fuhr. Phone: +41 44 633 7153

5 figures, 2 tables

**Summary**

The present work investigates environmental effects on the growth of fungal colonies of *P. vitreus* by using a lattice-free discrete modelling approach called FGM (Fuhr & Schubert, 2010, arXiv:1101.1747), in which hyphae *and* nutrients are considered as discrete structures.

A discrete modelling approach allows studying the underlying mechanistic rule concerning the basic architecture and dynamic of fungal networks on the scale of a single colony. By comparing simulations of the FGM to laboratory experiments of growing fungal colonies on malt extract agar we show that combined effect of temperature, pH and water activity on the radial growth rate fungal colony on a macroscopic scale may be explained by a power law for the growth costs of hyphal expansion on a microscopic scale. The information about the response of the fungal mycelium on a microscopic scale to environmental conditions is essential to simulate its behavior in complex structure substrates such as wood, where the impact of the fungus to the wood (i.e. the degradation of bordered pits or the creation of bore holes and cavities) changes the local environmental condition, e.g. the permeability of the substrate and therefore the water activity level in a colonized pore. A combination of diffusion processes of moisture into wood with the FGM may brighten the knowledge about the colonization strategy of *P. vitreus* and helps to optimize its growth behavior for biotechnological application such as bioincising.

## 1 Introduction

Microorganisms such as wood-decay fungi regulate their metabolism as respond to changing environmental conditions. Environmental factors thereby highly influenced the growth behavior and development of fungal mycelium in wood (Rayner and Boddy, 1988). Schubert et al. (2010) showed that the water activity, temperature and pH are the key factors for the growth behavior of the wood-decay fungi *Physisporinus vitreus* and *Neolentinus lepideus*. The knowledge about the influence of these abiotic factors to the microorganisms is of relevance for biotechnological applications of these wood-decay fungi in bioremediation (Majcherczyk and Hüttermann, 1998; Messner et al., 2002) or bioincising (Lehringer et al., 2009; Schubert and Schwarze, 2009) as well as for the morphogenesis of filamentous fungi in general.

The biotechnological process of bioincising is a method for enhancing the permeability of refractory wood by using wood-decay fungi such as *P. vitreus* in order to improve the liquid uptake (e.g. wood-modification agents or resins). The degradation of the bordered pits, which are valve-like connections between the wood pores and closed in Norway spruce heartwood, provides *P. vitreus* as a suitable agent for bioincising in its first stage of growth (Schwarze and Landmesser, 2000; Schwarze et al., 2006). In this growth phase the wood colonization process is accompanied only by slight mass losses and without reducing significantly the strength of the wood (Schwarze et al., 2006). The uptake of wood modification agents can be increased by a factor of 5 depending on the incubation conditions as shown by (Schubert and Schwarze, 2009). The clustered growth pattern of the fungus and its simultaneous growth pattern reported by (Lehringer et al., 2010) observe that the degradation of the pit membranes occurred simultaneously with an attack of the cell wall (i.e. bore holes, cavities and notches). The occurrence of such negative side effects may depend on influencing factors such as nutrient supply and moisture content of the substrate(Lehringer et al., 2010). Therefore the knowledge about the response of the fungus to combined environmental factors is of significance for the biotechnological applications since the optimal growth behavior (i.e. uniformity of wood colonization) is important for upscaling the bioincising process. Mathematical models in combination with laboratory experiments can illuminate the growth behavior of the fungus and may provide an optimization of biotechnological processes under defined conditions.

In order to model the response of a single fungal colony on combined effects of environmental factors semi-empirical models based on ordinary differential equations are widely used (Davidson, 2007; Skinner et al., 1994). The response surface (RS) methodology was successfully applied to model the growth of ascomycetes (Begoude et al., 2007; Panagou et al., 2003; Schubert et al., 2010; Schubert et al., 2009) as well as basidiomycete (Schubert et al.). Compared to RS radial basis function (RBF) neural network models, which are based on artificial neural networks, show a better performance to predict specific growth rate in relation to environmental factors (Panagou et al., 2007). By using RBF neural network Schubert et al. (2010) found that the growth of the fungi *P. vitreus* mainly depends on the water activity (0.950 – 0.998) and temperature (10-30 °C), while the pH (4-6) affects the growth rate to a lesser extent. Despite their success the class of models mentioned above is not able to explain the underlying mechanistic rules concerning the basic architecture and dynamic of fungal networks on the scale of a single colony. In particular, recent developments of discrete modelling approaches, in which hyphae are considered as discrete structures, successfully simulate fungal growth in heterogeneous (Boswell, 2008; Boswell et al., 2007) or even complex physically and chemically structured wood-like environments (Fuhr et al., 2010).

The present work investigates environmental effects on the growth of fungal colonies of *P. vitreus* by using a lattice-free discrete modelling approach, in which hyphae *and* nutrients are considered as discrete structures. Discrete models of fungal growth have been developed by many authors in the past (Bell, 1986; Cohen, 1967; Ermentrout and Edelstein-Keshet, 1993; Hutchinson et al., 1980; Kotov and Reshetnikov, 1990; Regalado et al., 1996). A review of discrete models is given by Boswell & Hopkins (Boswell and Hopkins, 2009). This class of models is not a substitution for classic modelling approaches such as RS or RBF, but rather a complement to investigate further aspects of hyphal growth on the single colony scale.

## 2 Materials and methods

### 2.1. Hyphal growth model

In order to analyze the growth of filamentous fungi in homogeneous environment, e.g. the growth of fungi on malt extract agar (MEA), we use a two-dimensional implementation of the hyphal growth model introduced by Fuhr et al. (2010). This hyphal growth model is a lattice-free discrete modelling approach for simulating fungal growth in physically and chemically complex structured environment.

The model considers the mycelium as an assemblage of nodes, edges and tips. The substrate is formed of point-like nutrient sources. The nodes are connected by edges and form the mycelium, whereby the nodes of the mycelium are restricted to the position of the nutrient points as shown in Fig. 1a. The evolution of the mycelium is driven by key processes, such as polarization, uptake- and concentration of nutrients, and transport of nutrients, determining the morphology of the fungal colony (see Fig. 1b). Details about the model description can be found in Fuhr et al. (2010).

In order to model fungal growth on a two-dimensional surface, i.e. the expansion of a colony on MEA, the substrate consists of Poisson distributed nutrient points with $v$ initial amount of nutrients. Therefore the probability $P$ to find $k$ nutrient points at a specific interval is given by

$$P(k) = \frac{\omega^k}{k!} e^{\omega}, \tag{1}$$

where $\omega > 0$ is a scale parameter of the distribution.

The simulation starts by placing $n_k^0$ starting nodes, called pellets, with an initial nutrient concentration $n_n^0$ on a circle with diameter $n_d^0$ millimeter in the centre of a two-dimensional surface of size $L \times L$. The initial nodes have tips with an orientation normal to the circle's surface. At every iteration step $m$ of the algorithm the mycelium is extended by one edge of length $l_i$ at node ($i$) if its nutrient concentration $f_i^{(m)}$ is larger than the growth costs $\Omega_i^{(m)}$. We assume that the growth costs of the fungus, which may be interpreted as the energy consumption of the metabolic activity of the fungus, increase with lower values of the water activity since the water activity is a measure for the available amount of water for an organism in its environment, e.g. the growth of fungus *P. vitreus* is inhibited below $a_w$ =

0.966 (Schubert et al., 2010). Dependent on the water activity level Schubert (Schubert et al., 2010) observed a maximal growth rate of the fungus at about 25°C. Below or above this optimum growth temperature the fungal colony expands at a lower rate, i.e. the growth costs are higher. The growth costs may be given by

$$\Omega_i^{(m)} = \left(a \cdot \frac{l_i^{(m)}}{\xi}\right)^b \cdot \upsilon, \qquad (2)$$

where $a$ and $b$ are unknown factors depend of the water activity and the temperature respectively. $\xi$ is the growth cut-off length (Fuhr et al., 2010). The scaling behavior of the unknown factors can be estimated by comparing the simulation with laboratory experiments.

## 2.2 Experimental design

In order to calibrate and verify the hyphal growth model we use light microscopy. The fungus was cultivated on a glass slide covered with MEA. The experiments were performed with a Zeiss 200M (10× 0.5NA Fluar objective) widefield microscopy (WFM) at room temperature and pH = 6. The growing colony was observed over a time span of 2.5 hours by taking single images of size 1024 × 1024 pixels (48 bit RGB color) with a resolution of approximately 0.78 µm per pixel. A 3 x 4 grid of single images built a mosaic image with a total size of 2077 × 4095 pixels. The stitching of the single images to a mosaic image was performed with the AxioVision software package. The mosaic images were taken with a camera (AxioCamMR3) at an interval of 15 minutes. We observe a radial growth rate of the fungal colony of about 2 mm per day, which corresponds to a water activity level of the environment of approximately 0.990 (Schubert et al., 2010).

## 2.3 Measures of the fungal colony

Filamentous fungi form a fractal, tree-like structure termed mycelium in order to explore their environment. By considering the mycelium as a network common network measures can be applied to study fungal colonies (Fricker et al., 2007).

Radial growth rate

The radial expansion of a fungal colony is often used as a measure for the metabolic activity of the fungus. We define the radius *R* of a fungal colony starting form a circular inoculum as

$$R_{95} = \min\left\{ r : \frac{\sum_{(i)\forall\,(d_i^2<r^2)}^{N_k} l_i}{\sum_{i=1}^{N_k} l_i} \geq 0.95 \right\}, \qquad (3)$$

where $N_k$ is the number of nodes, $l_i$ is the total length of mycelium associated with node *(i)* and $d_i$ is its distance from the centre of gravity of the inoculum. The radial growth rate $G_r$ is given by $G_r = dR_{95} / dt$.

Hyphal growth unit

The hyphal growth unit (HGU) is the average length of hypha associated with each tip of mycelium. First postulated by Plomley (1958), the HGU is defined as

$$HGU = \frac{\sum_{i=1}^{N_k} l_i}{N_t}, \qquad (3)$$

where $N_t$ is the number of active tips of the mycelium. The HGU depends on the environmental condition and is constant during unrestricted growth of the mycelium (Plomley, 1958; Trinci, 1974).

## 3 Results and discussion

The combined abiotic factors temperature, pH and water activity significantly influence the growth behavior and development of the wood-decay fungus *P. vitreus* as shown in Schubert et al. (2010). In order to investigate and discuss the underlying mechanism determining the growth behavior of the fungus we calibrate the FGM in a first step (Sec. 3.1) by comparing the microscopic growth pattern of the colony *in vitro* and *in silico*. The obtained parameter set can be use to estimate the parameters $a$ and $b$ of Eq. 2 by comparing the simulation with the laboratory experiments of Schubert et al. (2010) in Sec. 3.2. The knowledge about the scaling behavior of these parameters may be useful to simulate the growing fungus in complex structure environments such as wood, where the activity of the fungus (i.e. the degradation of bordered pits or the creation of bore holes and cavities) changes the local environmental conditions, e.g. the water activity level in a pore.

## 3.1 Growth pattern

Fig. 2a shows the growth front of *P. vitreus* measured with WFM and Fig. 2b the corresponding simulation by using the parameter set given in Tab. 1. The growth fronts of the colonies expand at room temperature with about $G_r$ = 2 millimeters per day, which corresponds to a water activity level of $a_w$ = 0.990. We measure the hyphal length of the mycelium, the number of active tips and the HGU of the colony front over a time span of 150 minutes within the box in Fig. 2. The results are presented in Tab. 2. We consider only the front of the colony because of the identification of the active tips in the core region of the fungal colony is difficult due to many overlapping layers of mycelium.

We observe that hyphae on the colony front growth much faster, than hyphae in the older part of the mycelium. The ratio of the velocities between these classes of hyphae is about 5/1. This effect is considered in the FGM by a lack of nutrients in core of the colony. However, the FGM does not consider an inhibitor, which restricts the growth of the hyphae in the core of the colony.

The automated identification of the hyphae measured with WFM is difficult and based on present data especially in the core region of the mycelium not possible. A fluorescent fungi can help to improve the contrast of the images and may an automated segmentation is possible.

3.2 Radial growth rate

Fig. 3 shows the radial growth rate measured at different temperatures and evaluated for four sets of water activity levels, i.e. the water activity levels are 0.998 (circle), 0.990 (triangle), 0.982 (square), 0.974 (diamond) and 0.970 (star). The solid lines with the empty symbols are the corresponding simulation of the FGM by fitting the model parameters *a* and *b* to the water activity $a_w$ and temperature *T* respectively. We use *a* = {0, 0.2, 1, 1.8, 6.4} for the water activities levels of $a_w$ = {1, 0.998, 0.990, 0.982, 0.974} and for the parameter *b* in the interval [0, 4] the scaling law $b(T) = 9 + 5 \cdot T$. Therefore we can estimate the model parameters *a* as function of the water activity $a_w$ as shown in Fig. 5 by

$$a(a_w) = 0.1985 \cdot \left(e^{134.5 \cdot (1-a_w)} - 1\right). \tag{4}$$

The model parameter *a* increases exponentially with decreasing water activity levels, while the parameter *b* depends linearly on the temperature.

The model shows, that the radial growth rate increases for higher water activity levels and decreases with lower temperature. For *a* = 0 the growth costs of the fungus vanish and we observe the maximal growth rate of 4.5 millimeters per day. Above approximately *a* = 6.4 there is no growth observable within one day, since the growth costs even for small hyphal lengths (i.e. $\frac{l_i^{(m)}}{\xi} \sim 0.16$) are higher than the available nutrients at a nutrient point. Between 10 and 25°C the model show a good agreement with quantitative (Fig. 3) and qualitative laboratory experiments (Figs. 2 and 4). Above 25°C the radial growth rate of the mycelium decreases and hyphal growth may be affected by heat shock proteins (Sienkiewicz et al., 1997). The effect of such proteins is not incorporated in our model.

## 4 Conclusion

We analyze the combined effect of temperature, pH and water activity on the radial growth rate of *Physisporinus vitreus* in a homogeneous environment by using a two-dimensional hyphal growth model considering hyphae and nutrients as discrete structures. The simulations show a good qualitative and quantitative agreement with experimental results.

We found that combined effect of temperature, pH and water activity on the radial growth rate fungal colony may be explained by growth costs for hyphal expansion on a microscopic scale described by the power law of Eq. 2. The presented model is limited to a temperature range between 5 and 25 °C and a water activity between 0.970 and 1. Above 25°C the radial growth rate of the mycelium decreases and hyphal growth may be affected by heat shock proteins. The effect of such proteins is not incorporated in our model.

Our results are of significance for biotechnological applications of *P. vitreus* in processes such as bioincising (Schubert and Schwarze, 2009) as well as for the study of filamentous fungi in general. For the process of bioincising, which is a biotechnological approach to improve the uptake of wood-modification substances by harnessing the selective degradation pattern of *P. vitreus*, the knowledge about the response of the fungus to combined abiotic factors is important for successfully designing an optimal incubation procedure. Therefore future studies will focus on modelling the growth of *P. vitreus* in heterogeneous and physically structured two-dimensional wood-like environments.

## 6 Acknowledgements

The authors gratefully acknowledge to J. Hehl for helpful discussions and express their gratitude to the Swiss National Science Foundation (SNF) No. 205321-121701 for its financial support. Imaging was performed on instruments of the ETH Light Microscopy Center.

**Figure captions**

Figure 1: Fungal growth model. (a) The fungal growth model (FGM) considers hyphae and nutrients as discrete objects. Hyphae are root-like branching structures of the filamentous fungus and represented in the model by edges and nodes. (b) Starting from initial nodes the growth of the fungal colony is determined by key processes such as polarization, branching, uptake and concentration of nutrients and transport. Details about the model construction can be found in (Fuhr et al., 2010).

Figure 2: Model calibration. In order to calibrate the FGM we compare the simulation to laboratory experiments of *P. vitreus* cultivated on MEA and observed over 150 minutes by using WFM. The qualitative comparison of the growth front (a) *in vitro* and (b) *in silico* by using the FGM with the parameter set given in Tab. 1 shows a good agreement. Quantitative comparisons of total hyphal length of the mycelium, the number of active tips and the HGU within the box are given in Tab. 2.

Figure 3: Effect of environment. Response of the fungus to different sets of water activity levels and temperatures at pH 5. The filled symbols are the laboratory experiments of(Schubert et al., 2010), i.e. the water activity levels are 0.998 (circle), 0.990 (triangle), 0.982 (square), 0.974 (diamond) and 0.970 (star). The solid lines with the empty symbols are the corresponding simulation of the FGM by using $a$ = [0.0, 0.2, 1.0, 1.8, 6.4] and values of $b$ in the interval [0.2, 4.0]. Each solid line represents the average over 10 realizations and the uncertainty of the data points is within the range of the symbols.

Figure 4: Growth pattern. (a) Morphologies of *P. vitreus* at water activity levels of 0.982, 0.990 and 0.999 from the left to the right at T = 20°C and pH = 6. (b) The corresponding simulations show the fungal colony after approximately 24 hours of growth using a = (0, 1, 1.8) and b = 2.

Figure 5: Parameter estimation. The model parameter $a$ increases exponentially with decreasing water activity levels.

**Table caption**

Table 1: Typical model parameters used throughout this work. We use the same notation as in (Fuhr et al., 2010).

Table 2: Quantitative comparison of the experiment and the FGM shown in Fig. 2. The evolution of the mycelium is measured *in vivo* over a time span of 150 minutes at a temperature of approximately 20 °C, pH = 6 and a water activity of 0.990. This experimental condition corresponds to the FGM by using the parameters in Tab. 1.

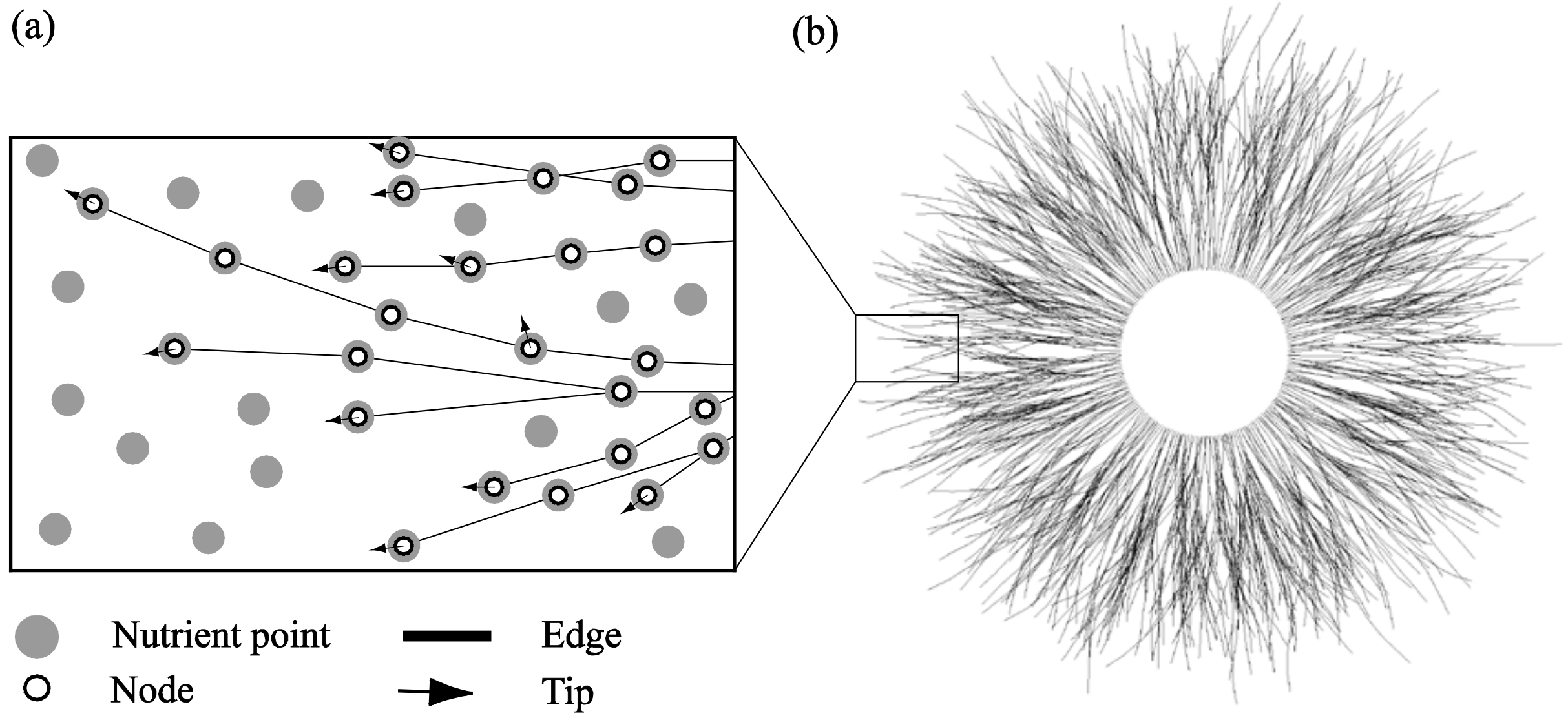

Fig.1: Abbildungen/Model/Model.eps
80 x 160 mm

Dieser Text ist in Times New Roman 12

**Dieser Text ist in Times New Roman 12 B**

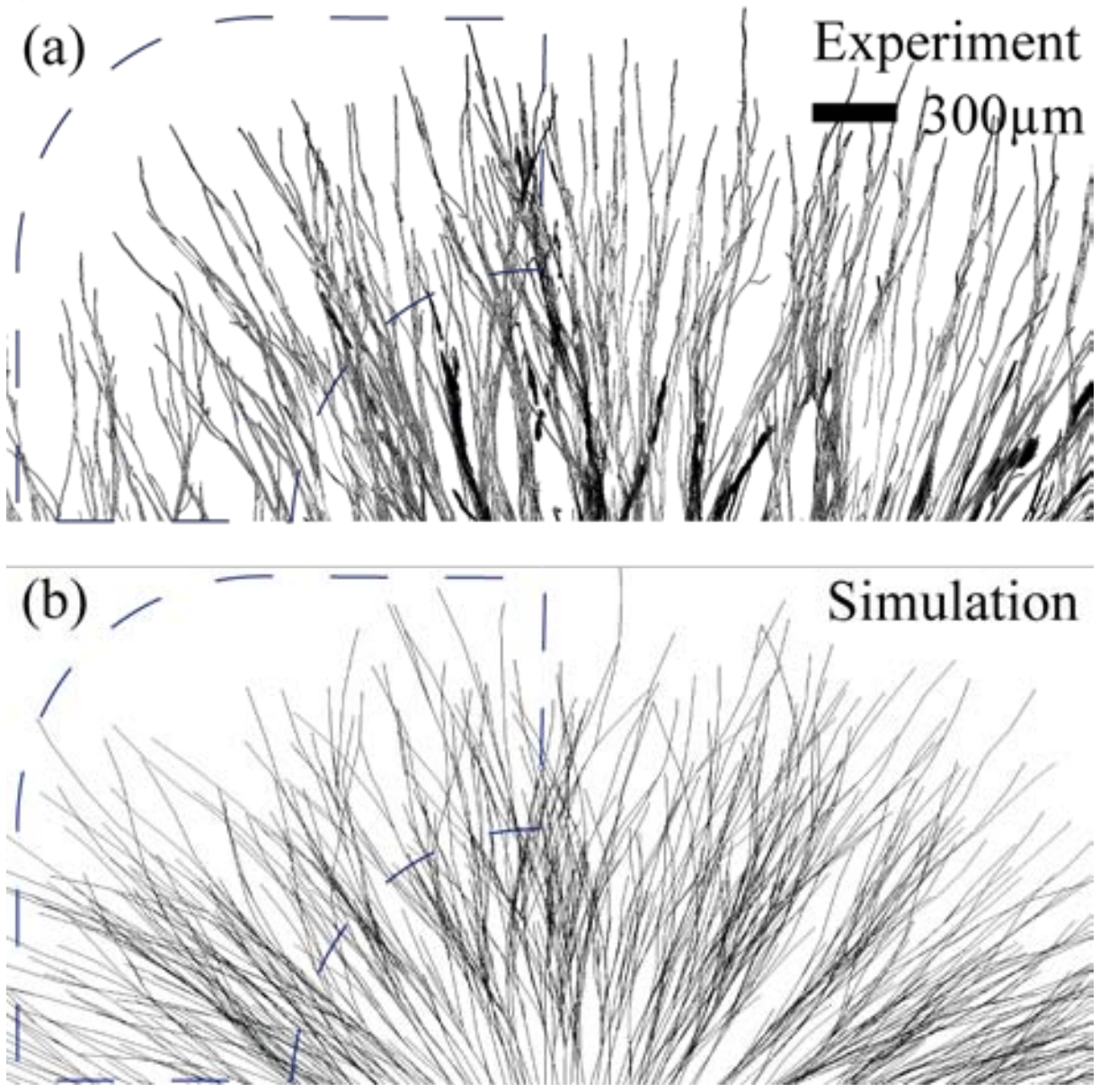

Fig. 2: Abbildungen/GrowthFront/GrowthFront_Box.png
80 x 80 mm

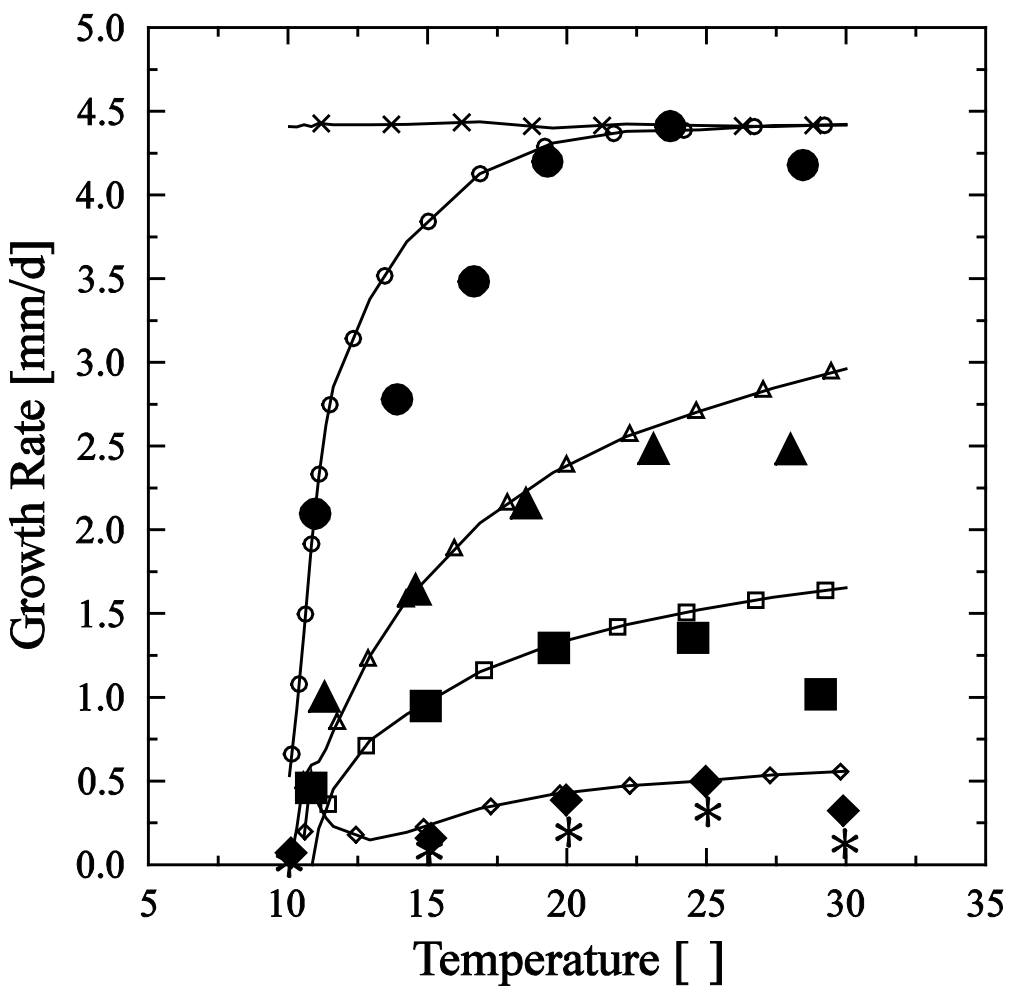

Fig. 3: /home/mfuhr/15_fungi3D/02_Models/20_2Dmorph/auswertung_22/PLOT_PAPER/
temp_aw_20_2Dmoph.eps

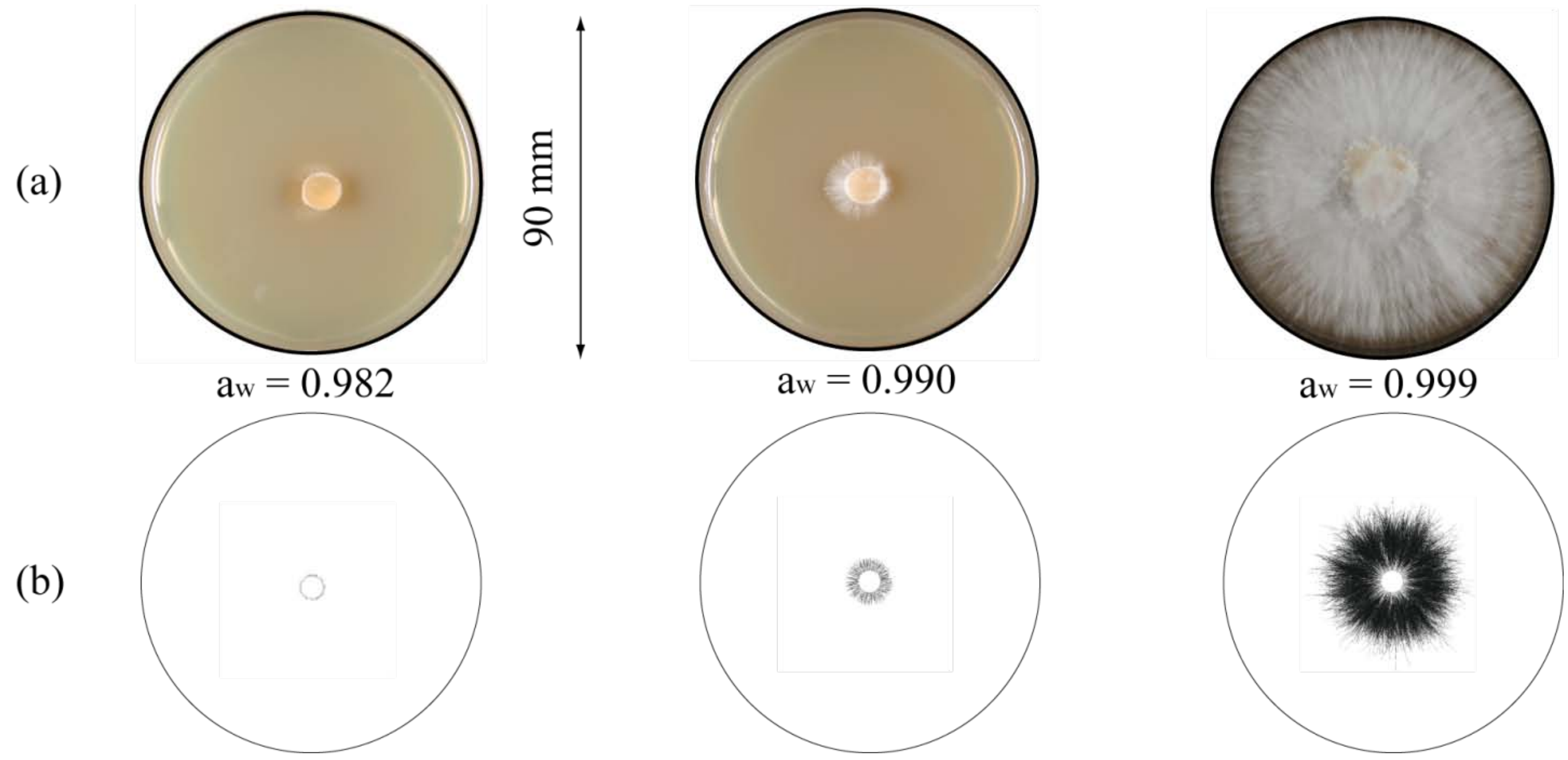

Fig. 4: Abbildungen/WaterActivity/WaterActivity1.png

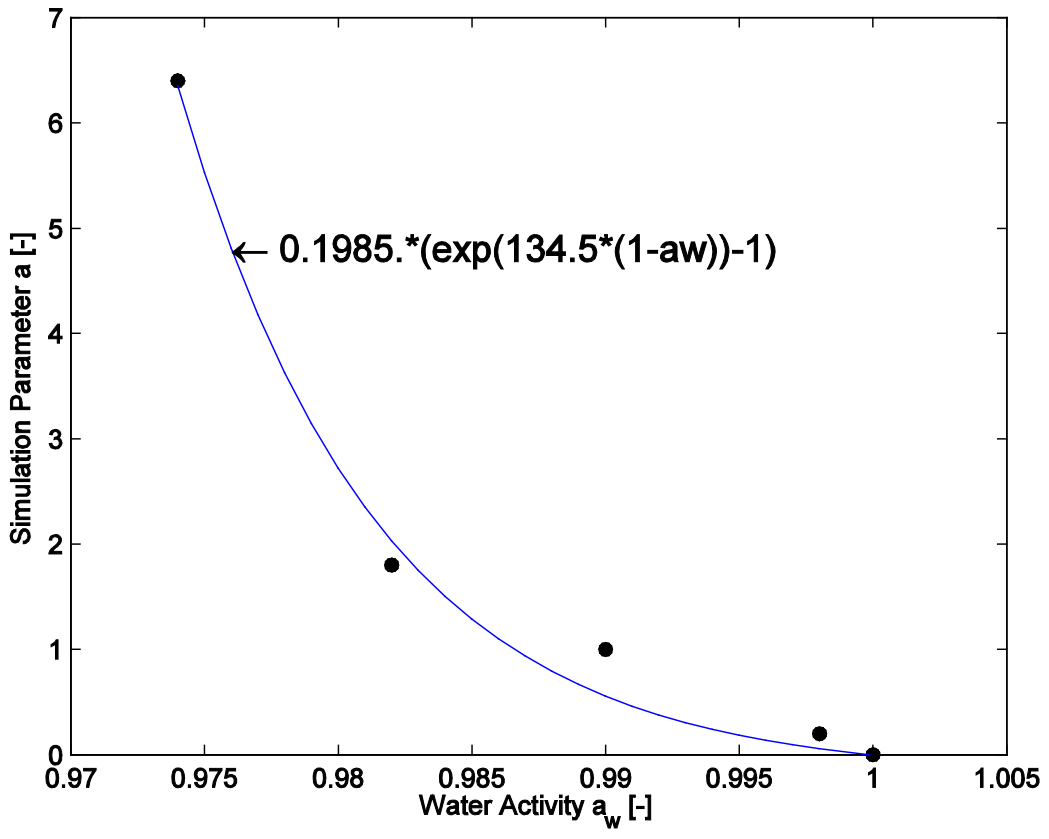

Fig. 5: /home/mfuhr/15_fungi3D/02_Models/20_2Dmorph/auswertung_22/FIT_a_b/FIT_a.eps

| | | | |
|---|---|---|---|
| **Substrate:** | | | |
| Box size L × L | $L$ | 90 | mm |
| Number of nutrient point | $N_p$ | 1.6 * 10^7 | - |
| Poissonian distribution (interval = 2*$\xi$) | $\omega$ | 2733 | - |
| | | | |
| **Fungus:** | | | |
| Mean hyphal growth rate | $\mu$ | 2.6 | mm/d |
| Mean edge length | $\lambda$ | 2/3*$\xi$ | mm |
| Growth cut-off length | $\xi$ | 10 | mm |
| Growth cut-off angle | $\theta$ | 0.44 | ° |
| Growth costs (Eq. 2) | $[a,b]$ | [1, 1.5] | - |
| Pit inital nutrient | $\nu$ | $4*10^{-13}$ | mol |
| Pit initial degradation rate | $\alpha_I$ | $\nu$/20 | mol |
| Pit degradation rate | $\alpha_C$ | 0.45*$\nu$ | mol/d |
| Apical branching threshold | $\beta_t$ | 0.6*$\nu$ | mol |
| Lateral branching threshold | $\beta_s$ | 0.35*$\nu$ | mol |
| | | | |
| **Simulation:** | | | |
| Initial number of pellet | $n_k^0$ | 200 | - |
| Initial number of tips | $n_s^0$ | 1 | - |
| Initial nutrient concentration | $n_n^0$ | 3/2*$\beta_t$ | mol |

Table 1

| | Symbol [unit] | Experiment | Model |
|---|---|---|---|
| | | | |
| Total hyphal length of mycelium | $L_m$ [mm] | 90 | 96 |
| Number of active tips | $N_t$ [-] | 210 (± 50) | 230 |
| Hyphal growth unit | HGU = $L_m$ / $N_t$ [µm] | 428 (± 90) | 417 |
| Radial growth rate | $G_R$ [mm/d] | 2 | 2.05 |

Table 2